\documentclass[%
 preprint,
 amsmath,amssymb,
 aps,
showkeys
]{revtex4-2}

\usepackage{graphicx}% Include figure files
\usepackage{dcolumn}% Align table columns on decimal point
\usepackage{bm}% bold math
\usepackage{subcaption}
\usepackage{mathtools}

\begin{document}

\title{Nuclear stability and the Fold Catastrophe}% Force line breaks with \\

\author{Samyak Jain \\
ORCID: 0009-0000-7455-782X}

\affiliation{Department of Physics, Indian Institute of Technology Bombay, 
Powai, Mumbai 400 076, India}%
\email{Corresponding author: samyakjain02@gmail.com}
\author{A. Bhagwat\\
ORCID: 0000-0002-3479-1301}
\email{Contributing author: ameeya@cbs.ac.in}
\affiliation{School of Physical Sciences, UM-DAE Centre for Excellence in Basic Sciences, University of Mumbai, Kalina Campus, Mumbai 400 098, India}%Authors' institution and/or address\\}%

\date{\today}% It is always \today, today,
             %  but any date may be explicitly aspecified

\begin{abstract}
A geometrical analysis of the stability of nuclei against deformations is presented. In particular, we use Catastrophe Theory to illustrate discontinuous changes in the behavior of nuclei with respect to deformations as one moves in the $N-Z$ space. We construct a minimalistic deformation model using the microscopic-macroscopic approach. A third-order phase transition is found in the liquid-drop model, which translates to a complete loss of stability (using the Fold catastrophe) when shell effects are included. The analysis is found to explain the instability of known fissile nuclei and also justify known decay chains of heavy nuclei. 
% \begin{description}
% Secondary publications and information retrieval purposes.
% You may use the \texttt{description} environment to structure your abstract;
% use the optional argument of the \verb+\item+ command to give the category of each item. 
% \end{description}
\end{abstract}

\maketitle
\newpage
%\tableofcontents

\section{Introduction} \label{introduction}
Developed in the 1970s by Ren\'e Thom, Catastrophe theory\cite{thom} is a geometrical framework that explains discontinuous changes in dynamical systems. Thom showed that any function of $r \le 5$ parameters can be mapped to one (and only one) of 11 known families of functions (catastrophes). One can thus study a wide range of systems by mapping their potential to one of these catastrophes. These catastrophes have been well-studied for discontinuous features, which can be mapped back to the system of interest to find where these changes occur. The proof of the theorem can be found in \cite{poston} and \cite{arnold}; we shall take it as a given.

The theory's application to quantum systems is trickier since one must use semi-classical methods to write a classical version of the quantum Hamiltonian. In particular, nuclei present a suitable testing ground for this since nuclear decays (especially fission) are well-studied discontinuous phenomena that would provide experimental verification for any analysis conducted. In this work, we consider the liquid-drop model and the shell model.

We note here that the expressions for deformation energy in most nuclear deformation models, including ours, are in the form of an infinite power series. Owing to the analytical nature of Catastrophe Theory, such a power series must be suitably truncated to obtain a polynomial. Thus, in this work we construct a minimalistic model and incorporate only the most important deformation parameter, and truncate our expressions so as to obtain reasonable agreement with known fission barriers, while still retaining the simplicity needed for an analytical treatment.

In Section \ref{liquid drop}, we outline the generalization of the liquid-drop model to deformed shapes as presented in \cite{shapeandshells}. In Section \ref{topt}, we then discuss the stability analysis of the model and find a third-order phase transition. In Section \ref{shell model}, we outline the shell model as presented in \cite{shapeandshells}, and how it is incorporated into the liquid-drop model as a correction term. In Section \ref{stability analysis}, we map the final deformation energy to the fold catastrophe and find a complete loss of stability as one moves in the $N-Z$ space, and see how this explains observed fissile nuclei and decay chains of heavy nuclei in Section \ref{verification}.

\section{The deformed liquid drop model}
\label{liquid drop}
We begin with the spherical liquid-drop model as presented in \cite{shapeandshells}: 
\begin{eqnarray}
    B &=& a_v A \left[ 1-k_v I^2 \right] 
    - a_s A^{2/3} \left[ 1- k_s I^2 \right] - a_C \frac{Z^2}{A^{1/3}} \label{spherical}
\end{eqnarray}
where $B$ is the binding energy of the nucleus, $a_v, a_s, a_C, k_v$ and $k_s$ are constants, $Z$ is the atomic number, $N$ is the neutron number and 
\begin{eqnarray}
    I = \frac{N-Z}{N+Z}
\end{eqnarray}
The values of the constants are obtained through best fits and have been stated in \cite{shapeandshells} and \cite{Myers} as
\begin{eqnarray}
  a_C = 0.7053, \quad a_s = 17.944 MeV, \quad k_s = 1.782
\end{eqnarray}
The parameter $k_v$ is in the range 1.5 - 2.0 \cite{shapeandshells}, however, the parameter is 
not relevant in the analysis reported here.

\begin{figure}[h]
\includegraphics[width=\linewidth]{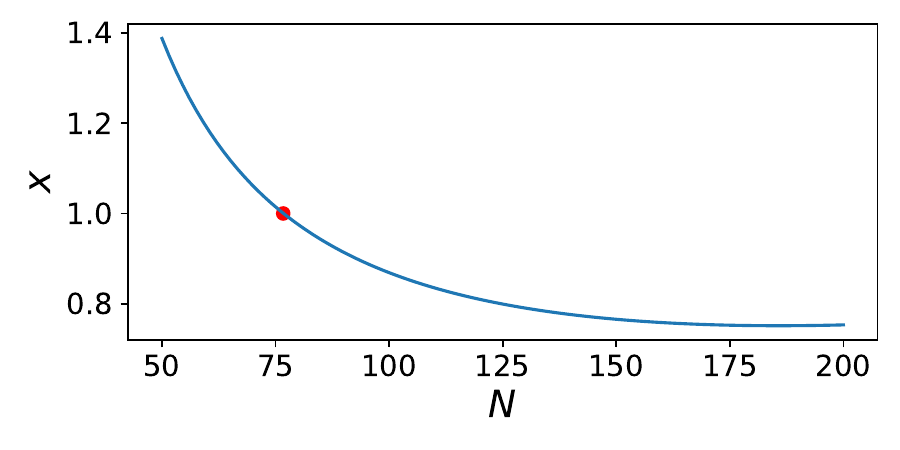}% Here is how to import EPS art
 \caption{$x$ as a function of $N$ for $Z = 92$. The point at which $x = 1$ is marked.}
 \label{x_92}
\end{figure}

The model has a volume term (proportional to the volume), a surface term (proportional to the surface area), and a coulomb term respectively. This model can be generalized to a deformed nucleus by simply accounting for the change in the surface area and coulomb interactions and rewriting $B$ as
\begin{eqnarray}
    B = a_s A^{2/3}\left( 1 - k_s I^2\right)  \frac{E_s(d)}{E_s^0}  + a_C \frac{Z^2}{A^{1/3}}\frac{E_C(d)}{E_C^0} 
    \label{deformed nucleus}
\end{eqnarray}
where we assume that any deformations conserve the volume of the nucleus, making the volume term irrelevant.
$E_C(d), E_s(d)$ are the surface and coulomb energies of the deformed nucleus, and $E_s^0, E_C^0$ are the surface and coulomb energies of the spherical nucleus
\begin{eqnarray}
    E_s^0 = a_s A^{2/3} (1 - k_sI^2) \label{surface spherical}
\end{eqnarray}
\begin{eqnarray}
    E_C^0 = a_C \frac{Z^2}{A^{1/3}}
\end{eqnarray}
The shape of a deformed nucleus can be parameterised using spherical harmonics as
\begin{eqnarray}
    R(\theta, \phi) = R_\alpha \left[ 1 + \sum_{\lambda = 1} ^ \infty \sum_{\mu = -\lambda} ^ \lambda \alpha_{\lambda \mu} Y_{\lambda \mu} (\theta, \phi)\right] \label{sph harmonic parameterization}
\end{eqnarray}
To simplify this, we assume that the deformations of the nucleus are symmetric about an axis (taken to be the $z$ axis). Thus, we can work with Legendre Polynomials instead of spherical harmonics; we only need to consider $\mu = 0$ in the summation. We define $\beta_\lambda = \alpha_{\lambda \mu}$ and use 
\begin{eqnarray}
    Y_{\lambda 0}(\theta, \phi) = \left(\frac{2\lambda + 1}{4\pi}\right) ^{1/2} P_\lambda(\cos(\theta))
\end{eqnarray}
We thus obtain
\begin{eqnarray}
    R(\theta) = R_\beta \left[ 1 + \left( \frac{2\lambda + 1}{4\pi} \right)^{1/2} \sum_{\lambda = 1} ^ \infty  \beta_\lambda P_\lambda (\theta)\right] \label{legendre poly expansion}
\end{eqnarray}
It turns out only $\beta_2$ (up to its third power) is relevant for small deformations (\cite{shapeandshells}). As shown in \cite{shapeandshells}, the Coulomb and the surface energies for small deformations can be calculated as
\begin{eqnarray}
    E_C = E_C^0 \left( 1 - \frac{1}{5}a_2^2 - \frac{4}{105}a_2 ^3 \right)
\end{eqnarray}

\begin{eqnarray}
    E_s = E_s^0 \left( 1 + \frac{2}{5}a_2^2 - \frac{4}{105}a_2^3\right)
\end{eqnarray}
where
\begin{eqnarray}
    a_2 = \sqrt{\frac{5}{4\pi}}\beta_2
\end{eqnarray}
Thus, the additional energy due to deformations is 
\begin{eqnarray}
    \Delta E &=& (E_s + E_C) - (E_s^0 + E_C^0) \nonumber
    \end{eqnarray}
    giving us
    \begin{eqnarray}
\Delta E = V_d = E_s^0\left( \frac{2}{5}(1-x)a_2^2 - \frac{4}{105}(1+2x)a_2^3 \right) \label{def energy}
\end{eqnarray}
where \begin{eqnarray}
    x = \frac{E_C^0}{2E_s^0} = \frac{a_C}{a_s}\frac{Z^2}{A}\frac{1}{1-k_sI^2}
    \label{fissility}
    \end{eqnarray}
x is the so-called fissility parameter; as \cite{shapeandshells} shows, it is a measure of how easy it is for a nucleus to undergo fission. We plot $x$ as a function of $N$ for $Z = 92$ in Fig.\ref{x_92}. 

We now examine the deformation energy $V_d$ (Eq. \ref{def energy}) for instabilities and phase transitions.

\section{The third-order phase transition}
\label{topt}
We note that both $x$ and $E_s^0$ are functions of $Z$ and $N$. Let us consider a fixed $Z$, for which we slowly vary $N$ from smaller to larger values. Since $Z$ is constant, we can say that $x$ and $E_s$ are functions of $N$. We consider $N$ to be a parameter, and treat $a_2$ as a coordinate. We assume that the ground state of the system dwells at the lowest energy minimum. We study how the system's energy (the energy the lowest energy minimum) varies with $N$.
\\
We have 
\begin{eqnarray}
    V_d(N) = E_s^0(N) \Bigl( a(N)a_2^3 + b(N)a_2^2  \Bigl)   \label{ld def energy} 
\end{eqnarray}
where 
\begin{eqnarray}
    a(N) &=& - \frac{4}{105} \left( 1+2x(N) \right) 
    \nonumber \\
      b(N) &=&  \frac{2}{5} \left( 1-x(N) \right)  \label{ab}
\end{eqnarray}
It can be easily shown from Eq.(\ref{surface spherical}) that $E_s(N)$ is a smooth and positive function of $N$ in the region of interest. Further, $a(N)<0$ for all $N>0$.

The fixed points of the system $a_2^0$ are given by
\begin{eqnarray}
    \frac{\partial V_d}{\partial a_2} = 0 \Rightarrow a_2^0 = 0, -\frac{2b}{3a} \label{fixed pts}
\end{eqnarray}
Further
\begin{eqnarray}
    \frac{\partial^2V}{\partial a_2^2}(0) &=& 2b 
    \nonumber \\
    \frac{\partial^2V}{\partial a_2^2}(-\frac{2b}{3a}) &=& -2b
\end{eqnarray}
For $b > 0$ the equilibrium at $a_2 = 0$ is stable and the equilibrium at $a_2 = \dfrac{-2b}{3a}$ is unstable. For $b<0$, the equilibrium at $a_2 = 0$ in unstable and the equilibrium at $a_2 = -\dfrac{2b}{3a}$ is stable.

Thus, at $b = 0$, the fixed points meet, and the stability of the two fixed points swap. For $b>0$ the system dwells at $a_2^0 = 0$. The energy of the system is
\begin{eqnarray}
    V_d(N, a_2^0 = 0) = 0 
\end{eqnarray}
This is completely independent of $N$. Thus
\begin{eqnarray}
    b > 0 \Rightarrow \frac{d^nV_d(a_2^0 = 0)}{dN^n} = 0 \quad \forall  n \in \mathbb{N}
\end{eqnarray}
Now, for $b < 0$, the system dwells at $a_2^0 = -\dfrac{2b}{3a}$. The energy of the system is given by 
\begin{eqnarray}
    V_d\left(N, a_2^0 = -\frac{2b}{3a}\right) = \frac{4b^3}{27a^2}E_s^0 
\end{eqnarray}
We now evaluate the derivatives of this energy with respect to $N$, at $b = 0$. We note that $a, b, E_s^0$ and their derivatives with respect to $N$ are smooth in our region of interest (where most observed nuclei are found in the $N-Z$ space). Also, as noted earlier, $a(N) <0 \quad \forall \quad N$. \\
\newline 
On computing the derivatives, one obtains multiple terms containing derivatives of $a, b, E_s^0$. However, since we are evaluating these derivatives at $b = 0$, most of these terms vanish. Using this, one finds easily that, at $b = 0$
\begin{eqnarray}
    \frac{dV_d(a_2^0)}{dN} = \frac{d^2V_d(a_2^0)}{dN^2} = 0 
\label{first two}
\end{eqnarray}
However 
\begin{eqnarray}
    \frac{d^3V_d(a_2^0)}{dN^3} = \left(\frac{8}{9} \frac{E_s^0}{27a^2}  \left(\frac{db}{dN}\right)^3\right)\bigg|_{b=0} \ne 0
    \label{third}
\end{eqnarray}

One can verify that $x$ is a decreasing function of $N$ (for a fixed $Z$) in the region of interest (see Fig.\ref{x_92}). From Eq.(\ref{ab}), we see that we cross $b = 0$ at $x = 1$. As N is increased from lower to larger values, $x$ keeps decreasing, and at some $N = N_0$, $x$ falls below 1. Before this point, $b < 0$, and the system's lowest minimum is always given by $a_2^0 = -\frac{2b}{3a}$ (note this is a function of $N)$. The first three derivatives of this energy are given by Eq.(\ref{first two}) and Eq.(\ref{third}).

At $x = 1$, this minimum stops changing with N and remains fixed at $a_2^0 = 0$. Now, all the derivatives of the system's energy turn to 0. From Eq.(\ref{first two}) and Eq.(\ref{third}), we see that the first two derivatives are continuous, while the third derivative is discontinuous, implying a third-order phase transition. 

One can find the $N = N_0$ at which this phase transition occurs using
\begin{eqnarray}
    x(N_0) = 1 = \frac{a_C}{a_s}\frac{Z^2}{A} \frac{1}{1 - k_s I^2} \label{n0}
\end{eqnarray}

As per this analysis, as one varies $N$ from lower to higher values (for a fixed $Z$), the nucleus simply goes from preferring a deformed shape to a spherical shape via a third-order phase transition. This in itself does not imply a catastrophic instability one would imagine is needed for fission. We thus now consider a more microscopic nuclear model: the shell model.

\section{The shell model and shell correction}
\label{shell model}
The liquid-drop model has had success in describing the `average' behavior of nuclei but fails to account for the existence of magic numbers (neutron and proton numbers for which a nucleus is unusually stable). This extra stability can naturally be attributed to the filling of energy levels. 

One considers each nucleon to be in an attractive potential caused by all the other nucleons. A typical choice of this potential (for an undeformed nucleus) is the modified-harmonic oscillator potential given by Nilsson (\cite{nilsson1} and \cite{shapeandshells})
\begin{eqnarray}
    V_{\text{MO}}(r) = \frac{1}{2}m\omega_0^2r^2 - \kappa\hbar \omega_0 \left( 2\textbf{l}.\textbf{s} + \mu (l^2 - \left\langle l^2\right\rangle_\mathcal{N}) \right) \label{Vmo}
\end{eqnarray}
where $\mathcal{N} = 2n+l$ is the principal quantum number. To account for Coulombic interactions and the Pauli exclusion principle, neutrons and protons are considered to experience separate potentials with different natural frequencies (\cite{shapeandshells})
\begin{eqnarray}
    \hbar \omega_0^{N,Z} = 41 A^{-\frac{1}{3}}\left( 1 \pm \frac{1}{3} \frac{N-Z}{A} \right)  
\end{eqnarray}
The energy levels for the modified harmonic oscillator potential are analytically obtained as (\cite{shapeandshells})
\begin{eqnarray}
    E_{\mathcal{N}ls} = \hbar \omega_0 \Bigl(  \mathcal{N} + \frac{3}{2} - \kappa \left[ 2s(l+\frac{1}{2}) - \frac{1}{2} \right] 
-\mu^\prime \left[ l(l+1) - \frac{\mathcal{N}(\mathcal{N}+3)}{2} \right] \Bigl) \label{energy levels}
\end{eqnarray}
To account for deformations, one simply considers deformations to the spherical potential (\cite{nilsson2} and \cite{shapeandshells}):
\begin{eqnarray}
    V =  \frac{1}{2}m\omega^2(\vec{\varepsilon})r^2   \Biggl(1 - \frac{2}{3} \varepsilon P_2(\cos{\theta_t}) 
    + \sum_{\lambda = 3,4...} \varepsilon_\lambda P_\lambda (\cos{\theta_t}) \Biggl)
    \label{shell def}
\end{eqnarray}
Here, the frequency is considered as a function of the deformations to account for the volume conservation of an equipotential surface, and the factor of $2/3$ allows for simpler calculations later. One can find the shifts in the energy levels due to the deformations using perturbation theory. Let these energy levels be 
$e_\nu (\varepsilon)$ (where $\nu$ iterates over each nucleon, not each energy level, so degenerate energy levels are counted multiple times). To obtain the total energy of the nucleus, one can't simply add up the individual nucleon energies because of overcounting of inter-nucleon interactions. Since the liquid-drop model accounts well for the average behavior of nuclei, one incorporates the shell effects as a correction to the liquid-drop model.

This correction procedure is described in \cite{shapeandshells}: since the `average' behavior of nuclei is reproduced well by the liquid-drop model, one adds the shell model's deformation energy to the liquid-drop model and subtracts an `averaged' shell model energy (so that the new corrected energy still describes the average behavior of nuclei correctly).

To compute this `averaged' shell model energy, consider the shell model's energy density 
\begin{eqnarray}
    g(e) = \frac{1}{\gamma} \sum_\nu \delta\left(\xi_\nu\right)
\end{eqnarray}
where $\gamma$ is an energy scale to make the argument of the Dirac-delta dimensionless, and $\xi_\nu = \dfrac{e-e_\nu}{\gamma}$. The total energy is then
\begin{eqnarray}
    E = \int eg(e)de = \sum_\nu e_\nu
\end{eqnarray}
irrespective of $\gamma$. To obtain an averaged energy density, one smears the energy levels by expanding out the Dirac-delta functions in a Hermite polynomial expansion
\begin{eqnarray}
    \delta(x) = \sum_{n=0}^\infty \frac{1}{\sqrt{\pi}}c_ke^{-x^2}H_{2k}(x) \label{dirac delta}
\end{eqnarray}
where
\begin{eqnarray}
    c_k = \frac{(-1)^k}{(2k)!! 2^k}
\end{eqnarray}
and truncating this expansion at some order $n$ to obtain
\begin{eqnarray}
    \tilde{g}(e) = \frac{1}{\gamma\sqrt{\pi}}\sum_\nu \sum_{k=0}^n c_ke^{-\xi_\nu^2}H_{2k}(\xi_\nu) \label{smeared density}
\end{eqnarray}

 Then, $\gamma$ decides the scale of this smearing, and is typically chosen as $\hbar \omega_0$. One chooses an $n$ so that the resulting energies are relatively independent of $\gamma$, and \cite{shapeandshells} shows this is observed at $n=3$.\\
 The difference of these energy densities is then
\begin{eqnarray}
    g(e) - \tilde{g}(e) = 
    \frac{1}{\gamma \sqrt{\pi}} \sum_\nu \sum_{k = n+1}^\infty c_k e^{-\xi_\nu^2}H_{2k}(\xi_\nu)
\end{eqnarray}

Then, as per the correction outlined earlier, the total energy is 
\begin{eqnarray}
     E  &=& E_{LD} + E_{corr} \nonumber\\
    &=& E_{LD} +  \sum_\nu \sum_{k=n+1}^\infty \int \frac{e\,c_k}{\gamma\sqrt{\pi}} e^{-\xi_\nu^2}H_{2k}(\xi_\nu) de 
\end{eqnarray}
where $E_{LD}$ is the liquid-drop energy. To find the energy due to deformation, we simply need to include the deformation dependence of the energy levels
\begin{eqnarray}
    V(\vec{\varepsilon}) = V_d(a_2) + \sum_\nu \frac{\partial E_{corr}}{\partial e_\nu} \Bigg|_{\vec{\varepsilon} = 0} \Delta e_\nu 
    \label{def potential}
\end{eqnarray}
where $V_d(a_2)$ is given by Eq.(\ref{ld def energy}). We shall later link the deformation parameters $\varepsilon$ and $a_2$.
\section{Stability analysis}
\label{stability analysis}
Since the shell model is being incorporated with the liquid-drop model, we consider only $\varepsilon$ in the expansion given by Eq.(\ref{shell def}):
\begin{eqnarray}
    V =  \frac{1}{2}m \omega^2(\vec{\varepsilon}) r^2 \left(1 - \frac{2}{3} \varepsilon P_2(\cos{\theta_t})\right)  \label{def potential 2nd order}
\end{eqnarray}
We can then write the perturbed (deformed) Hamiltonian as
\begin{eqnarray}
    H = H_0 + \varepsilon H^\prime \label{H_total}
\end{eqnarray}
where 
\begin{eqnarray}
    H^\prime = -\frac{2}{3}mw_0^2 r^2P_2(\cos{\theta})
\end{eqnarray}
The shifts in energy levels are then computed in \cite{shapeandshells} as
\begin{eqnarray}
    \Delta E = \left\langle \varepsilon H^\prime \right\rangle = \frac{1}{6} \varepsilon m \omega(\varepsilon)^2 \left\langle r^2 \right\rangle \frac{3m_j^2 - j(j+1)}{j(j+1)} 
    \label{delta_split}
\end{eqnarray}
where $j = l+s$  and a $2j+1$ degeneracy is broken. We assume that upon deformation, a nucleon will pick the $m_j$ state with the least energy. Thus, the energy shift for deformed energy levels is given by 
\begin{eqnarray}
        \Delta E = \left\langle \varepsilon H^\prime \right\rangle = \frac{1}{6} \varepsilon m\omega(\varepsilon)^2 \left\langle r^2 \right\rangle \frac{3m_{j0}^2 - j(j+1)}{j(j+1)} \label{perturbed_shell_energies}
\end{eqnarray}
where $m_{j0}$ is the least energetic $m_j$ state available while respecting the Pauli Exclusion principle. In the spherical case, each nucleon has unique quantum numbers $\mathcal{N}, l, j$. Thus, once a nucleon occupies any $m_j$ state, the set of quantum numbers $\mathcal{N}, l, j, m_j$ is still unique for any $m_j$. Thus, the least energetic state each nucleon can choose is
\begin{eqnarray}
    m_{j0} = 0
\end{eqnarray}

Finally, we impose volume conservation of the equipotential surface to obtain \cite{shapeandshells}
\begin{eqnarray}
    \omega(\varepsilon) = \omega_0\left( 1 + \frac{1}{9} \varepsilon^2 ...\right) \label{omega_def}
\end{eqnarray}
Plugging Eq.(\ref{omega_def}) and Eq.(\ref{perturbed_shell_energies}) in Eq.(\ref{def potential}), we obtain (up to third order in $\varepsilon$)
\begin{eqnarray}
    V(\vec{\varepsilon}) = V_d(\vec{\varepsilon}) + c(N,Z)\left( \varepsilon + \frac{2\varepsilon^3}{9} \right)
    \label{corrected pot}
\end{eqnarray}
where $c(N,Z)$ incorporates all the proportionality coefficients and is given by 
\begin{eqnarray}
    c(N,Z) = \sum_\nu \frac{1}{6}m\omega_0^2\langle r^2 \rangle \frac{3m_{j0}^2 - j(j+1)}{j(j+1)} \frac{\partial E_{corr}}{\partial e_\nu} \Bigg|_{\varepsilon = 0}
\end{eqnarray}
One can easily calculate $\langle r^2 \rangle$ to be \cite{shapeandshells}
\begin{eqnarray}
  \langle r^2 \rangle = \left( \mathcal{N}+\frac{3}{2}\right)\frac{\hbar}{m \omega_0}
\end{eqnarray}
leading to 
\begin{eqnarray}
    c(N,Z) = \sum_\nu \frac{1}{6}\hbar\omega_0\left(\mathcal{N} + \frac{3}{2}\right) \frac{3m_{j0}^2 - j(j+1)}{j(j+1)} \frac{\partial E_{corr}}{\partial e_\nu} \Bigg|_{\varepsilon = 0}
\end{eqnarray}
To link the deformation $\varepsilon$ with the deformation $a_2$ considered in the liquid-drop model, we simply consider an equipotential surface for the deformed potential (Eq.(\ref{def potential 2nd order})), given by
\begin{eqnarray}
    r(\theta) = r_0 \left( 1 + \frac{2\varepsilon}{3}P_2(\cos{\theta}) \right)
\end{eqnarray}
Comparing this to the spatial deformations considered in the liquid-drop model (Eq.(\ref{legendre poly expansion}), we immediately see 
\begin{eqnarray}
    \varepsilon = \frac{3}{2}\sqrt{\frac{5}{4\pi}}\beta_2 =\frac{3}{2} a_2 \label{varepsilon 2}
\end{eqnarray}
Substituting this and $V_d(a_2)$ in Eq.(\ref{corrected pot}), we obtain 
\begin{eqnarray}
    V(a_2) = A a_2^3 + Ba_2^2 + Ca_2 \label{macro micro def energy}
\end{eqnarray}
where
\begin{eqnarray*}
A = E_s^0(N) a(N,Z) + \frac{3c(N,Z)}{4}
\end{eqnarray*}
\begin{eqnarray*}
 B = E_s^0(N,Z) b(N,Z)
\end{eqnarray*}
\begin{eqnarray}
    C = \frac{3}{2}c(N,Z)
\end{eqnarray}
We now appeal to Catastrophe theory. An obvious choice of a catastrophe to map $V(a_2$) is the fold catastrophe:
\begin{eqnarray}
    V_t(u) = u^3 + tu
\end{eqnarray}
where $t$ is a parameter and $u$ is a coordinate. Notably, the fold catastrophe has a pitchfork bifurcation at $t = 0$. For $t<0$, the potential has a maximum and a minimum, which meet and annihilate at $t = 0$. For $t>0$, no stable point exists in the system, leading to a drastic change in the system's behavior.\\
One can easily check that $Ax^3 + Bx^2 + Cx$ can be mapped to $u^3 + tu + s$ via the mapping
\begin{eqnarray}
    x = \frac{u}{A^{\frac{1}{3}}} + p
\end{eqnarray}
where 
\begin{eqnarray}
    p = \frac{-B}{3A}, \quad t = \frac{3AC-B^2}{3A^\frac{4}{3}}, \quad s = \frac{2B^3 - 9ABC}{27A^2}
\end{eqnarray}
Here, $s$ is a additive term with no coordinate dependence and can be safely discarded. Under this mapping, one observes the loss of stability at $t>0$, which translates to 
\begin{eqnarray}
    t > 0 \Rightarrow \frac{3AC-B^2}{A^{4/3}} > 0
\end{eqnarray}

Defining this expression as $f(N,Z)$, we get
\begin{eqnarray}
      f(N,Z) &=& \left[ E_s^0(N) a(N,Z) + \frac{3}{4}c(N,Z) \right]^{-4/3} \times
      \nonumber \\ &&
      \Biggl( \frac{9}{2}\left[\frac{3}{4}c(N,Z) +  E_s^0(N) a(N,Z)  \right] c(N,Z) \Big[E_s^0(N,Z)b(N,Z) \Big]^2\Biggl) \quad > \quad 0\label{ineq_macro_micro}
\end{eqnarray}

% \vspace{-0.8cm}

\section{Experimental verification}
\label{verification}
% Let us define the expression in the above inequality as $f(N, Z)$:
% \begin{eqnarray}
%   f(N,Z)=\left[ E_s^0(N) a(N,Z) + \frac{3c(N,Z)}{8} \right]^{-4/3}\Biggl( \Big[\frac{3c(N,Z)}{8} \nonumber\\ + E_s^0(N) a(N,Z)  \Big] \frac{9}{2}c(N,Z)
%       \nonumber\\
%     - \left[ E_s^0(N,Z)b(N,Z) \right]^2\Biggl)
%     \label{expression}
% \end{eqnarray}

\begin{table}[b]
\caption{\label{fission barriers}
Calculated and experimental fission barriers\textsuperscript{\cite{RIPL2}} for heavy nuclei. We obtain reasonable agreement in spite of us considering only a single deformation parameter.}
\begin{ruledtabular}
\begin{tabular}{cccccccc}
 Nucleus& Calculated (MeV)& Experimental ($E_b$) (MeV)& Experimental ($E_a$) (MeV) \\
\hline Po208 & 15.29 & 19.9\\
Po209 & 15.55 & 21.1 \\
Po210 & 15.80 & 21.2 \\
Th230 & 9.29 & 6.80 & 6.10 \\
Th231 & 9.44 & 6.70 & 6.00 \\
Th232 & 9.57 & 6.70 & 5.80 \\
U234 & 7.28 & 5.50 & 4.80 \\
U235 & 7.40 & 6.00 & 5.25 \\
U238 & 7.75 & 5.50 & 6.30 \\

\end{tabular}
\end{ruledtabular}

\end{table}

To compare with observables, first let we compare the fission barriers of heavy nuclei in our model to their experimental values \textsuperscript{\cite{RIPL2}}. This comparison is tabulated in Table.\ref{fission barriers}. Because of the minimalistic nature of the model, we only observe one barrier. However, we obtain reasonable agreement in spite of our minimalistic approach.

More intuitively, we plot $f(N,Z)$ in the $N-Z$ space. The values of the shell model parameters $\kappa$ and $\mu$ for various $\mathcal{N}$ levels are taken from \cite{shapeandshells} and are shown in Table \ref{table}.

\begin{table}[b]
\caption{\label{table}
$\mu$ and $\kappa$ values for various ${\mathcal{N}} = 2n+l+1$ levels}
\begin{ruledtabular}
\begin{tabular}{cccccccc}
 ${\mathcal{N}} = 2n+l$&$\mu$ &$\kappa $\\
\hline 2 & 0 & 0.8\\
3&0.0263&0.075\\
4,5,6,7&0.024&0.06

\end{tabular}
\end{ruledtabular}

\end{table}

\begin{figure}[h]
\includegraphics[width=0.9\linewidth]{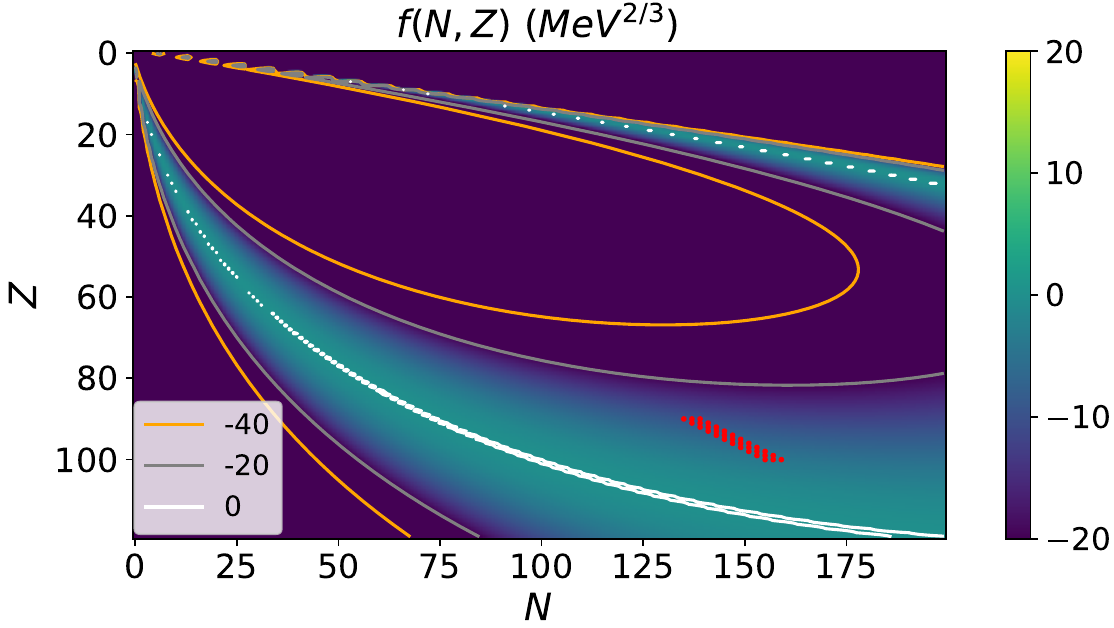}
 \caption{$f(N,Z)$ is plotted in the $N-Z$ plane. A band-like structure is seen in which $f(N,Z)$ is positive or close to 0, indicating a loss of stability. Various contours are added, and nuclei satisfying Ronen's fissile rule are scattered in red. The contour of $f(N, Z) = 0$ is marked in white; these nuclei favor spontaneous deformations from a spherical shape.} 
 \label{macro_micro_contours_fig}
\end{figure}

\begin{figure*}[htbp]
\includegraphics[scale=0.55]{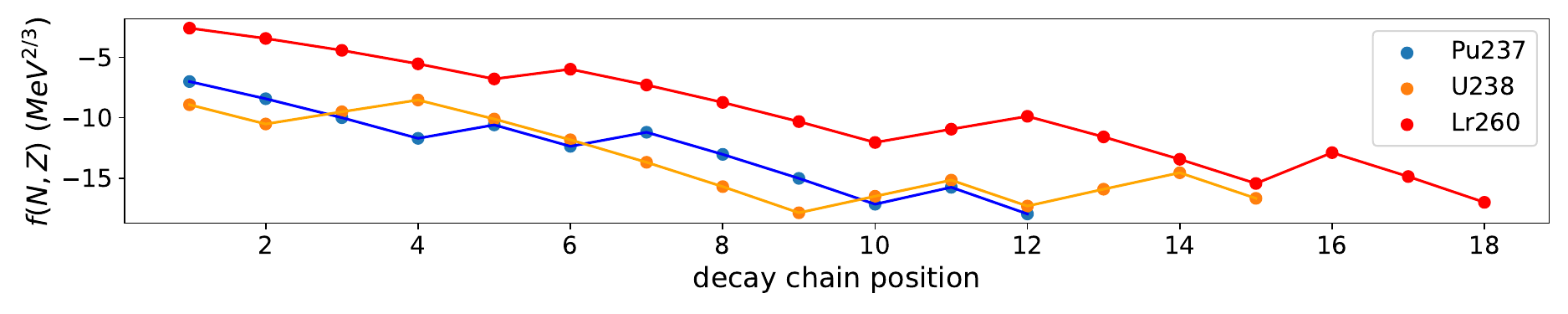}
 \caption{\label{decays}Evolution of $f(N,Z)$ in various decay chains is shown. Note the jumps at each $\beta$ decay. Decay chains: 
 U238$\xrightarrow{\alpha}$Th234$\xrightarrow{\beta}$Pa234$\xrightarrow{\beta}$U234$\xrightarrow{\alpha}$Th230$\xrightarrow{\alpha}$Ra226$\xrightarrow{\alpha}$Rn222$\xrightarrow{\alpha}$Po218$\xrightarrow{\alpha}$Pb214$\xrightarrow{\beta}$Bi214$\xrightarrow{\beta}$Po214
$\xrightarrow{\alpha}$Pb210$\xrightarrow{\beta}$Bi210$\xrightarrow{\beta}$Po210$\xrightarrow{\alpha}$Pb206 \newline
Pu237$\xrightarrow{\alpha}$U233$\xrightarrow{\alpha}$Th229$\xrightarrow{\alpha}$Ra225$\xrightarrow{\beta}$Ac225$\xrightarrow{\alpha}$Fr221$\xrightarrow{\beta}$Ra221$\xrightarrow{\alpha}$Rn217$\xrightarrow{\alpha}$Po213,Pb209$\xrightarrow{\beta}$Bi209$\xrightarrow{\alpha}$Tl205 \newline
Lr260$\xrightarrow{\alpha}$Md256$\xrightarrow{\alpha}$Es252$\xrightarrow{\alpha}$Bk248$\xrightarrow{\alpha}$Am244$\xrightarrow{\beta}$Cm244$\xrightarrow{\alpha}$Pu240$\xrightarrow{\alpha}$U236$\xrightarrow{\alpha}$Th232$\xrightarrow{\alpha}$Ra228$\xrightarrow{\beta}$Ac228$\xrightarrow{\beta}$Th228
\newline $\xrightarrow{\alpha}$Ra224$\xrightarrow{\alpha}$Rn220$\xrightarrow{\alpha}$Po216$\xrightarrow{2\beta}$Rn216$\xrightarrow{\alpha}$Po212$\xrightarrow{\alpha}$Pb208} 
 \label{Pa231}
\end{figure*}

$c(N,Z)$ is symbolically evaluated in python. We plot $f(N, Z)$ as a heat map in the $N-Z$ space, leading to Fig.(\ref{macro_micro_contours_fig}).
We observe a band-like structure, with nuclei in the interior of the band having $f(N, Z)<0$ and being stable and those near the periphery having higher values of $f(N, Z)$. We note here that any nuclei with a positive $f(N, Z)$ are inherently unstable, and we would not expect to see them at all. Instead, we can use $f(N, Z)$ as a measure of how unstable a nucleus is, with higher values implying less stability. We also add some contours and mark nuclei satisfying Ronen's fissile rule (\cite{ronen}): nuclei satisfying
\begin{eqnarray}
    90\leq Z \leq100,\quad 2Z - N = 43 \pm 2
\end{eqnarray}
are fissile. We see that all the nuclei lie close to the band periphery, where there is a sudden sharp increase in $f(N, Z)$ (implying a sudden increase in instability). Another test is to see how $f(N, Z)$ varies across decay chains. We show this for various decay chains in Fig.\ref{decays}.

We observe a drop in $f(N, Z)$ (implying an increase in stability) as each decay progresses, consistent with our analysis. Notably, all $\alpha$ decays are seen to decrease $f(N,Z)$. However, we observe that each $\beta$ decay increases $f(N, Z)$, with a drop again in consequent $\alpha$ decays. Understanding this feature requires further investigation, but this naturally paints a picture of a $\beta$ decay being an `intermediate' step that allows the decay chain to end at a more stable nucleus (as compared to the one it would have ended at without a $\beta$ decay), at the expense of some temporary instability in the form of an increase in $f(N, Z)$.

\vspace{-0.4cm}

\section{Conclusion}
\label{conclusion}
We have presented a geometric analysis of the stability of nuclei against deformations using Catastrophe theory, by constructing a minimalistic microscopic-macroscopic deformation model. The model accounts for a single deformation parameter to preserve the analytical nature of this work, while also retaining enough complexity to account for known fission barriers. We find a third-order phase transition in the liquid-drop model as one moves in the $N-Z$ space, in which the nucleus simply goes from preferring a spherical shape to preferring a deformed shape with no characteristic loss of stability. Upon incorporating shell effects, this translates from a phase transition to a complete loss of stability characteristic of the Fold catastrophe. 

Experimentally, our analysis is seen to explain the instability of fissile nuclei and also explain various decay chains of heavy nuclei, especially $\alpha$ decays. We also find good agreement with experimentally observed fission barriers.

\section{Acknowledgements}
\vspace{-0.2cm}
We would like to thank Vikram Rentala and Kumar Rao for their valuable insights.

\textbf{Funding:} No funding was received for conducting this study.
\newline
\textbf{Conflict of interest:} The authors of this work declare that they have no conflicts of interest.

% \section{Data Availability}
% Data sharing is not applicable to this article as no data sets were generated or analyzed during the current study.

\bibliographystyle{unsrt}
\bibliography{main}
\end{document}